\documentclass[prl,floatfix,superscriptaddress,twocolumn]{revtex4}
\usepackage[final]{pdfpages}
\usepackage{graphicx}
\usepackage{color}
\usepackage[normalem]{ulem}
\usepackage{braket}

\newcommand{\be}{\begin{equation}}
\newcommand{\ee}{\end{equation}}
\newcommand{\ba}{\begin{eqnarray}}
\newcommand{\ea}{\end{eqnarray}}

\begin{document}
\title{Entanglement in a quantum annealing processor}
\author{T.~Lanting\footnote[1]{Electronic address: tlanting@dwavesys.com}}
\author{A.~J.~Przybysz}
\author{A.~Yu.~Smirnov}
\affiliation{D-Wave Systems Inc., 3033 Beta Avenue, Burnaby BC
Canada V5G 4M9}
\author{F. M. Spedalieri}
\affiliation{Information Sciences Institute, University of Southern California, Los Angeles CA USA 90089}
\affiliation{Center for Quantum Information Science and Technology, University of Southern California}
\author{M.~H.~Amin}
\affiliation{D-Wave Systems Inc., 3033 Beta Avenue, Burnaby BC Canada V5G 4M9}
\affiliation{Department of Physics, Simon Fraser University, Burnaby, British Columbia, Canada V5A 1S6}
\author{A.~J.~Berkley}
\author{R.~Harris}
\author{F.~Altomare}
\affiliation{D-Wave Systems Inc., 3033 Beta Avenue, Burnaby BC Canada V5G 4M9}
\author{S.~Boixo\footnote[2]{currently at Google, 340 Main St, Venice, California 90291}}
\affiliation{Information Sciences Institute, University of Southern California, Los Angeles CA USA 90089}
\author{P.~Bunyk}
\affiliation{D-Wave Systems Inc., 3033 Beta Avenue, Burnaby BC Canada V5G 4M9}
\author{N.~Dickson\footnote[3]{currently at Side Effects Software, 1401-123 Front Street West, Toronto, Ontario, Canada}}
\author{C.~Enderud}
\author{J.~P.~Hilton}
\author{E.~Hoskinson}
\author{M.~W.~Johnson}
\author{E.~Ladizinsky}
\author{N.~Ladizinsky}
\author{R.~Neufeld}
\author{T.~Oh}
\author{I.~Perminov}
\author{C.~Rich}
\author{M.~C.~Thom}
\author{E.~Tolkacheva}
\affiliation{D-Wave Systems Inc., 3033 Beta Avenue, Burnaby BC Canada V5G 4M9}
\author{S.~Uchaikin}
\affiliation{D-Wave Systems Inc., 3033 Beta Avenue, Burnaby BC Canada V5G 4M9}
\affiliation{National Research Tomsk Polytechnic University, 30 Lenin Avenue, Tomsk, 634050, Russia}
\author{A.~B.~Wilson}
\author{G.~Rose}
\affiliation{D-Wave Systems Inc., 3033 Beta Avenue, Burnaby BC Canada V5G 4M9}

\begin{abstract} Entanglement lies at the core of quantum algorithms designed to solve problems
that are intractable by classical approaches.
One such algorithm, quantum annealing (QA), provides a promising
path to a practical quantum processor. We have built a series of
scalable QA processors consisting of networks of manufactured
interacting spins (qubits). Here, we use qubit tunneling
spectroscopy to measure the energy eigenspectrum of two- and
eight-qubit systems within one such processor,
demonstrating quantum coherence in these systems. We present
experimental evidence that, during a critical portion of QA, the
qubits become entangled and that entanglement persists even as these
systems reach equilibrium with a thermal environment. Our results
provide an encouraging sign that QA is a viable technology for
large-scale quantum computing.
\end{abstract}

\maketitle

\section{I. Introduction}

The last decade has been exciting for the field of quantum
computation. A wide range of physical implementations of
architectures that promise to harness quantum mechanics to perform
computation have been
studied~\cite{mariantoni-science-2011,lucero-factoring-2012,reed-error-correction-2012}.
Scaling these architectures to build practical processors with many
millions to billions of qubits will be
challenging~\cite{fowler-pra-2012,metodi-2005}. A simpler
architecture, designed to implement a single quantum algorithm such
as quantum annealing (QA), provides a more practical approach in the
near-term~\cite{NaturePaper,DicksonNatCom13}. However, one of the
main features that makes such an architecture scalable, namely a
limited number of low bandwidth external control
lines~\cite{harris-2010}, prohibits many typical characterization
measurements used in studying prototype universal quantum
computers~\cite{Blatt08,Monz11,ansmann,Neeley10,dicarlo-nature-2010,Berkley03}.
These constraints make it challenging to experimentally determine
whether a scalable QA architecture, one that is inevitably coupled
to a thermal environment, is capable of generating entangled
states~\cite{vidal2003,Guhne09,GHZ,wootters1998}. A demonstration of
entanglement is considered to be a critical milestone for any
approach to building a quantum computing technology. Herein, we
demonstrate an experimental method to detect entanglement in
subsections of a quantum annealing processor to address this
fundamental question.

\section{II. Quantum annealing}

QA is designed to find the low energy configurations of systems of
interacting spins. A wide variety of optimization problems naturally
map onto this physical system
~\cite{Finnila,Kadowaki,Farhi,Santoro}. A QA algorithm is described
by a time-dependent Hamiltonian for a set of $N$ spins, $i = 1,
\ldots, N$,

\begin{equation}
 {\cal H}_S(s) = {\cal E}(s)\,{\cal H}_P -\frac{1}{2}\, \sum_{i} \Delta(s)\, \sigma^x_i \label{HS},
 \end{equation}
 where the dimensionless ${\cal H}_P$ is
 \begin{equation}
 {\cal H}_P = -\sum_{i}h_i\sigma^z_i + \sum_{i<j} J_{ij}\sigma^z_i\sigma^z_j
 \label{HP}
 \end{equation}
and $\sigma^{x,z}_i$ are Pauli matrices for the $i$th spin. The
energy scales $\Delta$ and ${\cal E}$ are the transverse and
longitudinal energies of the spins, respectively, and the biases
$h_i$ and couplings $J_{ij}$ encode a particular optimization
problem. The time-dependent variation of $\Delta$ and ${\cal E}$ is
parameterized by $s\equiv t/t_f$ with time $t\in [0,t_f]$ and total
run (anneal) time $t_f$. QA is performed by first setting $\Delta
\gg {\cal E}$, which results in a ground state into which the spins
can be easily initialized~\cite{NaturePaper}. Then $\Delta$ is
reduced and ${\cal E}$ is increased until ${\cal E}\gg \Delta$. At
this point, the system Hamiltonian is dominated by ${\cal H}_P$,
which represents the encoded optimization problem. At the end of the
evolution a ground state of ${\cal H}_P$ represents the lowest
energy configuration for the problem Hamiltonian and thus a solution
to the optimization problem.

\begin{figure}[h!]
 \includegraphics[width=7.5cm]{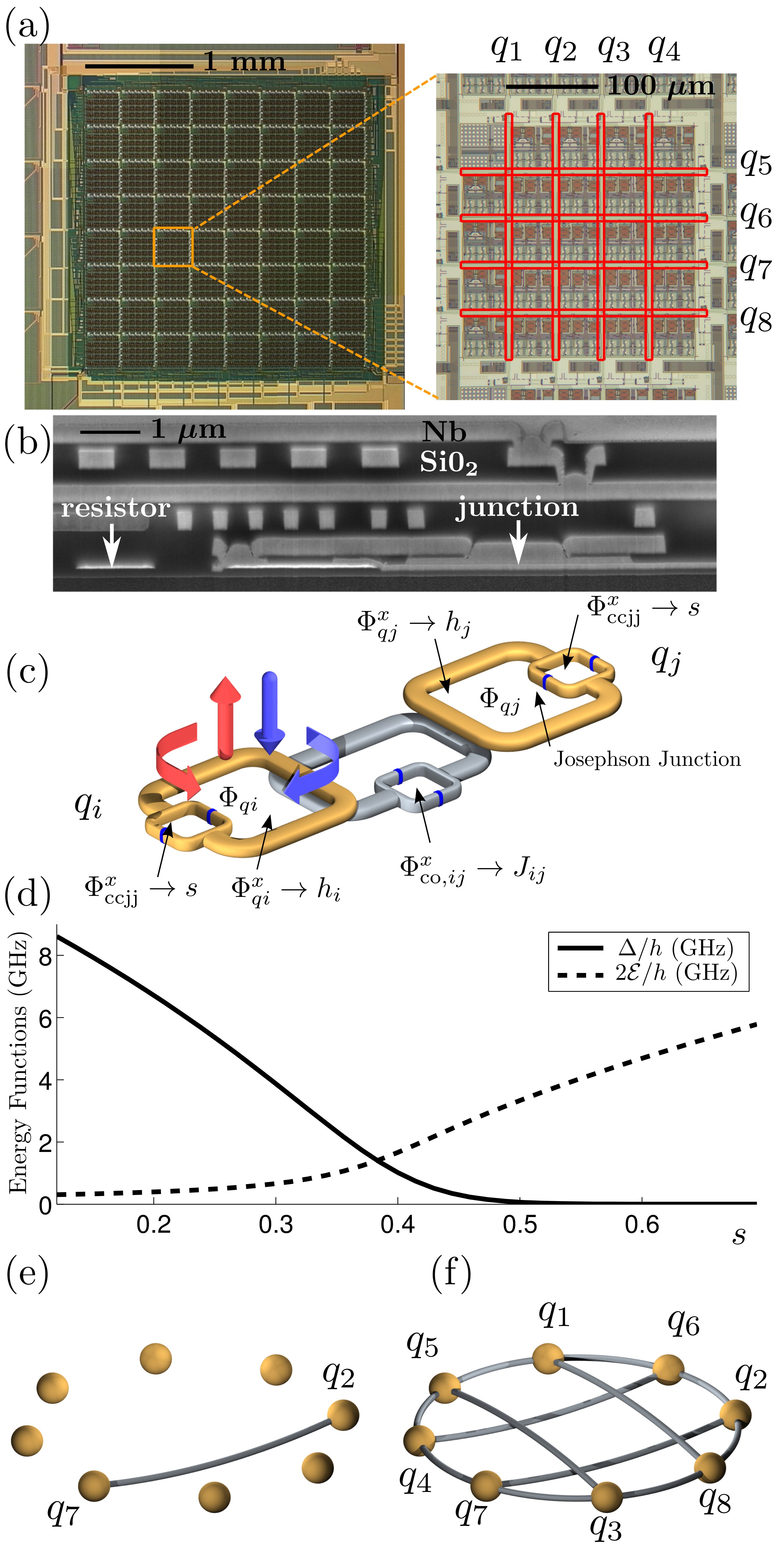}
\caption{{\bf (a)} Photograph of the QA processor
  used in this study. We report measurements performed on the
  eight-qubit unit cell indicated. The bodies of the qubits are
  extended loops of Nb wiring (highlighted with red
  rectangles). Inter-qubit couplers are located at the intersections
  of the qubit bodies. {\bf (b)} Electron micrograph showing the
  cross-section of a typical portion of the processor circuitry
  (described in more detail in Appendix A). {\bf (c)} Schematic
  diagram of a pair of coupled superconducting flux qubits with
  external control biases $\Phi^x_{qi}$ and $\Phi^x_{\rm ccjj}$ and
  with flux through the body of the $i$th qubit denoted as
  $\Phi_{qi}$. An inductive coupling between the qubits is tuned with
  the bias $\Phi^x_{{\rm co},ij}$. {\bf (d)} Energy scales $\Delta(s)$
  and ${\cal E}(s)$ in Hamiltonian (\ref{HS}) calculated from an
  rf-SQUID (Superconducting Quantum Interference Device) model based
  on the median of independently measured device parameters for these
  eight qubits. See Appendix A for more details. {\bf (e),(f)} The two
  and eight-qubit systems studied were programmed to have the
  topologies shown. Qubits are represented as gold spheres and
  inter-qubit couplers, set to $J = -2.5$, are represented as silver
  lines. \label{qubitschematic}}
\end{figure}

\begin{figure}[h!]
 \includegraphics[trim=0cm 0cm 0cm
   0cm,clip,width=8cm]{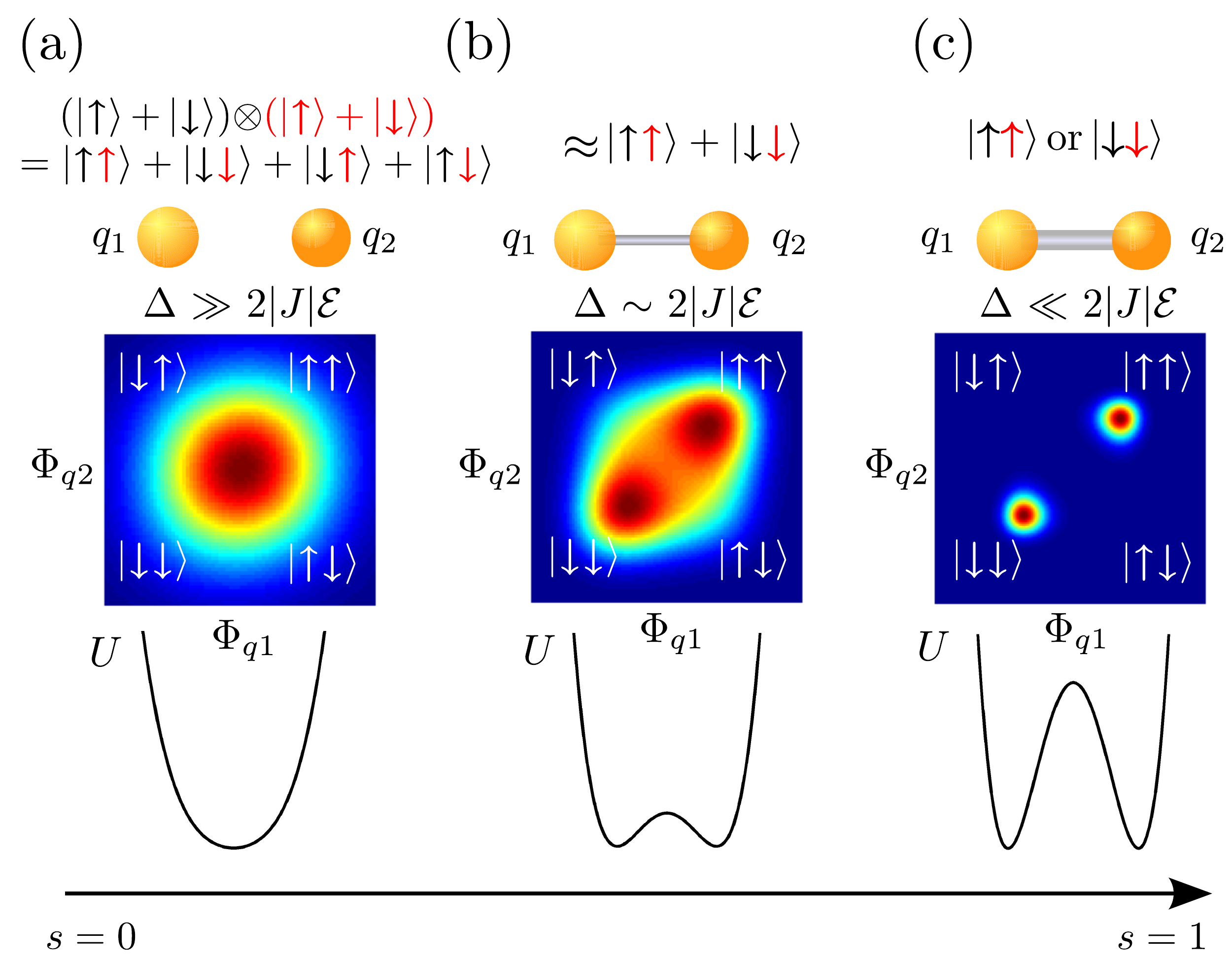}
\caption{\label{FIG:entanglement-cartoon} An illustration of
  entanglement between two qubits during QA with $h_i=0$ and $J < 0$.
  We plot calculations of the two-qubit ground state wave function
  modulus squared in the basis of $\Phi_{q1}$ and $\Phi_{q2}$, the
  flux through the bodies of $q_1$ and $q_2$, respectively. The color
  scale encodes the probability density with red corresponding to high
  probability density and blue corresponding to low probability
  density.  We used Hamiltonian~(\ref{HS}) and the energies in
  Fig~\ref{qubitschematic}d for the calculation. The four quadrants
  represent the four possible states of the two-qubit system in the
  computation basis.  We also plot the single qubit potential energy
  ($U$ versus $\Phi_{q1}$) calculated from measured device
  parameters. {\bf (a)} At $s$ = 0 ($\Delta \gg 2 |J| {\cal E}\sim
  0)$, the qubits weakly interact and are each in their ground state
  $\frac{1}{\sqrt{2}}(\ket{\uparrow}+\ket{\downarrow})$, which is
  delocalized in the computation basis. The wavefunction shows no
  correlation between $q_1$ and $q_2$ and therefore their wavefunctions are
  separable.  {\bf (b)} At intermediate $s$ ($\Delta \sim 2 |J_{ij}|
  {\cal E}$), the qubits are entangled. The state of one qubit is not
  separable from the state of the other, as the ground state of the
  system is approximately $\ket{+}\equiv (\ket{\uparrow\uparrow } +
  \ket{\downarrow\downarrow})/\sqrt{2}$. A clear correlation is seen
  between $q_1$ and $q_2$. {\bf (c)} As $s \rightarrow 1$, $\Delta \ll
  2 |J_{ij}| {\cal E}$ and the ground state of the system approaches
  $\ket{+}$. However, the energy gap $g$ between the ground state
  ($\ket{+}$) and the first excited state ($\ket{-}$) is closing. When
  the qubits are coupled to a bath with temperature $T$ and $g <
  k_BT$, the system is in a mixed state of $\ket{+}$ and $\ket{-}$ and
  entanglement is extinguished.}
\end{figure}

\section{ III. Quantum annealing processor}

We have built a processor that implements ${\cal H}_S$ using
superconducting flux qubits as effective
spins~\cite{NaturePaper,DicksonNatCom13,perdomo-folding-2012,boixo-annealing}.
Figure~\ref{qubitschematic}a shows a photograph of the processor.
Figure \ref{qubitschematic}c shows the circuit schematic of a pair of
flux qubits with the magnetic flux controls $\Phi_{qi}^x$ and
$\Phi_{\rm ccjj}^x$. The annealing parameter $s$ is controlled with
the global bias $\Phi_{\rm ccjj}^x(t)$ (see Appendix A for the mapping
between $s$ and $\Phi_{\rm ccjj}^x$ and a description of how $\Phi^x_{qi}$ is
provided for each qubit). The strength and sign of the inductive
coupling between pairs of qubits is controlled with magnetic flux
$\Phi_{{\rm co},ij}^x$ that is provided by an individual on-chip
digital-to-analog converter for each coupler~\cite{harris-2010}. The
parameters $h_i$ and $J_{ij}$ are thus {\em in situ} tunable, thereby
allowing the encoding of a vast number of problems. The time-dependent
energy scales $\Delta(s)$ and ${\cal E}(s)$ are calculated from
measured qubit parameters and plotted in
Fig.~\ref{qubitschematic}d. We calibrated and corrected the individual
flux qubit parameters in our processor to ensure that every qubit had
a close to identical $\Delta$ and ${\cal E}$ (the energy gap $\Delta$
is balanced to better than $8\%$ between qubits and ${\cal E}$ to
better than $5\%$). See Appendix A for measurements of these energy
scales. The interqubit couplers were calibrated as described in
Ref.~\cite{harris-cjc-2009}. The processor studied here was mounted on
the mixing chamber of a dilution refrigerator held at temperature $T =
12.5$ mK.

\section{IV. Ferromagnetically Coupled Instances}
The experiments reported herein focused on one of the eight-qubit unit
cells of the larger QA processor as indicated in
Fig.~\ref{qubitschematic}a. The unit cell was isolated by setting all
couplings outside of that subsection to $J_{ij}=0$ for all
experiments. We then posed specific ${\cal H}_P$ instances with strong
ferromagnetic (FM) coupling $J_{ij}=-2.5$ and $h_i=0$ to that unit
cell as illustrated in Figs.~\ref{qubitschematic}e and f. These
configurations produced coupled two- and eight-qubit systems,
respectively. Hamiltonian~(\ref{HS}) describes the behaviour of these
systems during QA.

Typical observations of entanglement in the quantum computing
literature involve applying interactions between qubits, removing
these interactions, and then performing measurements. Such an approach
is well suited to gate-model architectures (e.g.~Ref.~\cite{ansmann}).
During QA, however, the interaction between qubits is determined by
the particular instance of ${\cal H}_P$, in this case a strongly
ferromagnetic instance, and cannot be removed. In this way, systems of
qubits undergoing QA have much more in common with condensed matter
systems, such as quantum magnets, for which interactions cannot be
turned off. Indeed, a growing body of recent theoretical and
experimental work suggests that entanglement plays a central role in
many of the macroscopic properties of condensed matter
systems~\cite{amico2008,wang02,ghosh2003,Vertesi06,Brukner06,Rappoport07,Christensen07}.
Here we introduce other approaches to quantifying entanglement that
are suited to QA processors. We establish experimentally that the two-
and eight-qubit systems, comprising macroscopic superconducting flux
qubits coupled to a thermal bath at 12.5 mK, become entangled during
the QA algorithm.

To illustrate the evolution of the ground state of these instances
during QA, a sequence of wave functions for the ground state of the
two-qubit system is shown in Fig.~\ref{FIG:entanglement-cartoon}. A
similar sequence could be envisioned for the eight-qubit system. We
consider these systems subject to zero biases, $h_i=0$. For small $s$,
$\Delta \gg 2 |J_{ij}| {\cal E}$, and the ground state of the system
can be expressed as a product of the ground states of the individual
qubits: $\otimes_{i=1}^N
\frac{1}{\sqrt{2}}(\ket{\uparrow}_i+\ket{\downarrow}_i)$ where $N=2,8$
(see Fig.~\ref{FIG:entanglement-cartoon}a). For intermediate $s$,
$\Delta \lesssim 2 |J_{ij}| {\cal E}$, and the ground and first
excited states of the processor are approximately the delocalized
superpositions $\ket{\pm} \equiv (\ket{\uparrow ...  \uparrow } \pm
\ket{\downarrow ... \downarrow})/\sqrt{2}$
(Fig.~\ref{FIG:entanglement-cartoon}b). The state $\ket{+}$ is the
maximally entangled Bell (or GHZ, for eight qubits)
state~\cite{GHZ}. As $s\rightarrow 1$, the energy gap $g$ between the
ground and first excited states approaches $g \equiv (E_{2}-E_{1})
\propto \Delta(s)^N/(2|J_{ij}|{\cal E}(s))^{N-1}$ and vanishes as
$\Delta(s) \rightarrow 0$ (Fig.~\ref{FIG:entanglement-cartoon}c). At
some point late in the evolution, $g$ becomes less than $k_BT$, where
$T$ characterizes the temperature of the thermal environment to which
the qubits are coupled. At this point, we expect the system to evolve
into a mixed state of $\ket{+}$ and $\ket{-}$ and the entanglement
will vanish with $g$ for sufficiently long thermalization times. At
the end of QA, $s=1$, $\Delta \sim 0$, and Hamiltonian~(\ref{HS})
predicts two degenerate and localized ground states, namely the FM
ordered states $\ket{\uparrow ...\uparrow}$ and $\ket{\downarrow
  ... \downarrow}$.

\section{V. Measurements}

\begin{figure*}
\centering \setlength\fboxsep{0pt} \setlength\fboxrule{0pt}
\fbox{\includegraphics[trim=0cm 0cm 0cm
    0.cm,clip,width=18.cm]{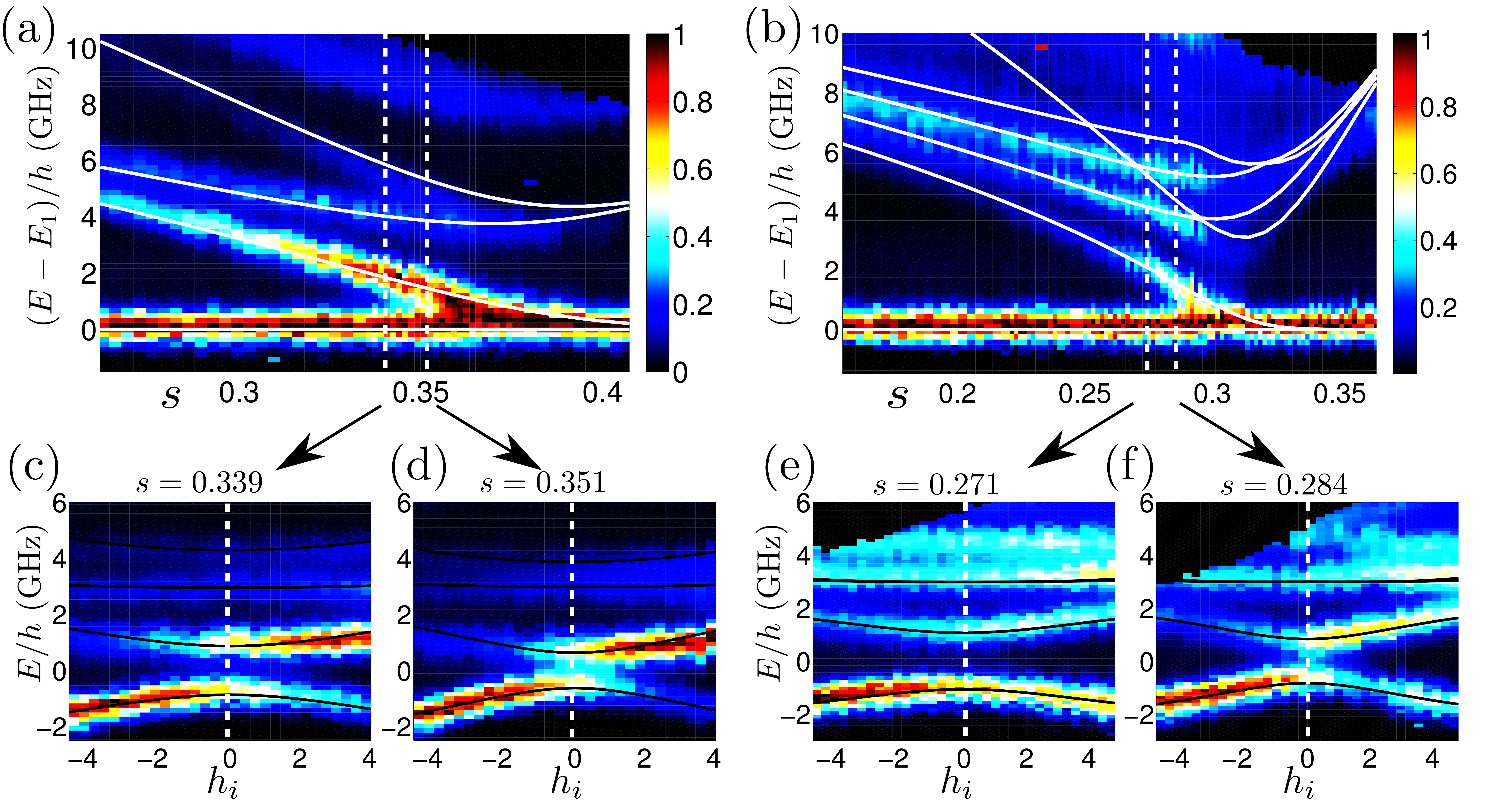}}
\caption{\label{FIG:wide-spectrum-summary} Spectroscopic data for two-
  and eight-qubit systems plotted in false colour (colour indicates
  normalized qubit tunnel spectroscopy rates). A non-zero measurement
  (false colour) indicates the presence of an eigenstate of the probed
  system at a given energy (ordinate) and $s$ (abscissa). Panel {\bf
    (a)} shows the measured eigenspectrum for the two-qubit system as
  a function of $s$.  Panel {\bf (b)} shows a similar set of
  measurements for the eight-qubit system.  The ground state energy
  $E_1$ has been subtracted from the data to aid in visualization. The
  solid curves indicate the theoretical expectations for the energy
  eigenvalues using independently calibrated qubit parameters and
  Hamiltonian (\ref{HS}). We emphasize that the solid curves are not a
  fit, but rather a prediction based on Hamiltonian~(\ref{HS}) and
  measurements of $\Delta$ and ${\cal E}$. The slight differences
  between the high-energy spectrum prediction and measurements are due
  to the additional states in the rf-SQUID flux qubits. A full
  rf-SQUID model that is in agreement with the measured high energy
  spectrum is explored in the Supplementary Information. Panel {\bf
    (c)} and {\bf (d)} show measured eigenspectra of the two-qubit
  system vs. $h_1 = h_2\equiv h_i$ for two values of annealing
  parameter $s$, $s=0.339$ and $s=0.351$ from left to right,
  respectively. Notice the avoided crossing at $h_i=0$. Panel {\bf
    (e)} and {\bf (f)} show analogous measured eigenspectra for the
  eight-qubit system with (with $h_1=\ldots=h_8 \equiv h_i$).  Because
  the eight-qubit gap closes earlier in QA for this system, we show
  measurements for smaller $s$, $s=0.271$ and $s=0.284$ from left to
  right, respectively.}
\end{figure*}

In order to experimentally verify the change in spectral gap in the
two- and eight-qubit systems during QA, we used qubit tunneling
spectroscopy (QTS) as described in more detail in
Ref.~\cite{berkley2012} and Appendix B. QTS allows us to measure the
eigenspectrum and level occupation of a system during QA by coupling
an additional probe qubit to the system. We performed QTS on the two-
and eight-qubit systems shown in Figs.~\ref{qubitschematic}e and
f. Figures~\ref{FIG:wide-spectrum-summary}a and b show the measured
energy eigenspectrum for the two- and eight-qubit systems,
respectively, as a function of $s$. The measurements are initial
tunneling rates of the probe qubit, normalized by the maximum observed
tunneling rate. Peaks in the measured tunneling rate map the energy
eigenstates of the system under study~\cite{berkley2012}. As the
system evolves (increasing $s$),~$\Delta(s)$ in Hamiltonian (\ref{HS})
decreases and the gap between ground and first excited states closes.
The spectroscopy data in Fig.~\ref{FIG:wide-spectrum-summary}a reveal
two higher energy eigenstates. We observe a similar group of higher
energy excited states for the eight-qubit system in
Fig.~\ref{FIG:wide-spectrum-summary}b. Note that $g$ closes earlier in
the QA algorithm for the eight-qubit system as compared to the
two-qubit system. In all of the panels of
Fig.~\ref{FIG:wide-spectrum-summary}, solid curves indicate the
theoretical energy levels predicted by Hamiltonian~(\ref{HS}) using
the measured $\Delta(s)$ and ${\cal E}(s)$. The agreement between the
experimentally obtained spectrum and the theoretical spectrum is good.

The data presented in Figs.~\ref{FIG:wide-spectrum-summary}a
and~\ref{FIG:wide-spectrum-summary}b indicate that the spectral gap
between ground and first excited state decreases monotonically with
$s$ when all $h_i=0$. Under these bias conditions, these systems
possess $Z_2$ symmetry between the states
$\ket{\uparrow\ldots\uparrow}$ and
$\ket{\downarrow\ldots\downarrow}$. The degeneracy between these
states is lifted by finite $\Delta(s)$. To explicitly demonstrate that
the spectral gap at $h_i=0$ is due to the avoided crossing of
$\ket{\uparrow\ldots\uparrow}$ and $\ket{\downarrow\ldots\downarrow}$,
we have performed QTS at fixed $s$ as a function of a ``diagnostic''
bias $h_i\neq 0$ that was uniformly applied to all qubits, thus
sweeping the systems through degeneracy at $h_i = 0$. As a result,
either the state $\ket{\uparrow...\uparrow}$ or $\ket{\downarrow
  ...\downarrow}$ becomes energetically favored, depending upon the
sign of $h_i$. Hamiltonian~(\ref{HS}) predicts an avoided crossing, as
a function of $h_i$, between the ground and first excited states at
degeneracy, where $h_i$ = 0, with a minimum energy gap $g$. The
presence of such an avoided crossing is a signature of ground-state
entanglement~\cite{Berkley03,WitPaper}. For large gaps, $g > k_BT$,
there is persistent entanglement at equilibrium (see
Refs.~\cite{wootters1998,amico2008,ghosh2003,Vertesi06,Rappoport07}
and the Supplementary Information).

We experimentally verified the existence of avoided crossings at
multiple values of $s$ in both the two- and eight-qubit systems by
using QTS across a range of biases $h_i \in \{-4, 4\}$. In
Fig.~\ref{FIG:wide-spectrum-summary}c we show the measured spectrum of
the two-qubit system at $s = 0.339$ up to an energy of 6 GHz for a
range of bias $h_i$.  The ground states at the far left and far right
of the spectrum are the localized states $\ket{\downarrow\downarrow}$
and $\ket{\uparrow\uparrow}$, respectively. At $h_i=0$, we observe an
avoided crossing between these two states. We measure an energy gap
$g$ at zero bias, $h_i=0$, between the ground state and the first
excited state, $g/h = 1.75 \pm 0.08$ GHz by fitting a Gaussian profile
to the tunneling rate data at these two lowest energy levels and
subtracting the centroids.  Here $h$ (without any subscript) is the
Planck constant.  Figure~\ref{FIG:wide-spectrum-summary}d shows the
two-qubit spectrum later in the QA algorithm, at $s = 0.351$. The
energy gap has decreased to $g/h = 1.21\pm 0.06$ GHz. Note that the
error estimates for the energy gaps are derived from the uncertainty
in extracting the centroids from the rate data. We discuss the actual
source of the underlying Gaussian widths (the observed level
broadening) below. For both the two- and eight-qubit system, we
confirmed that the expectation values of $\sigma_z$ for all devices
change sign as the system moves through the avoided crossing (see
Figs. 1-3 of the Supplementary Information and \cite{WitPaper})

Figures~\ref{FIG:wide-spectrum-summary}e and f show similar
measurements of the spectrum of eight coupled qubits at $s = 0.271$
and $s = 0.284$ for a range of biases $h_i$. Again, we observe an
avoided crossing at $h_i = 0$. The measured energy gaps at $s= 0.271$
and $0.284$ are $g/h = 2.2 \pm 0.08$ GHz and $g/h = 1.66 \pm 0.06$
GHz, respectively.  Although the eight qubit gaps in
Figs.~\ref{FIG:wide-spectrum-summary}e and f are close to the two
qubit gaps in Figs.~\ref{FIG:wide-spectrum-summary}c and d, they are
measured at quite different values of the annealing parameter $s$. As
expected, the eight-qubit gap is closing earlier in the QA algorithm
as compared to the two-qubit gap.  The solid curves in Figs.~c-f
indicate the theoretical energy levels predicted by
Hamiltonian~(\ref{HS}) and measurements of $\Delta(s)$ and ${\cal
  E}(s)$. Again, the agreement between the experimentally obtained
spectra and the theoretical spectra is good.

For the early and intermediate parts of QA, the energy gap $g$ is
larger than temperature, $g \gg k_BT$, for both the two- and
eight-qubit systems. We expect that if we hold the systems at these
$s$, then the only eigenstate with significant occupation will be the
ground state. We confirmed this by using QTS in the limit of long
tunneling times to probe the occupation fractions. Details are
provided in Appendix C. Figures~\ref{FIG:entanglement-degree}a and b
show the measured occupation fractions of the ground and first excited
states as a function of $s$ for both the two- and eight-qubit
systems. The solid curves show the equilibrium Boltzmann predictions
for $T = 12.5$ mK and are in good agreement with the data.

\begin{figure*}
\centering \setlength\fboxsep{0pt} \setlength\fboxrule{0pt}
\fbox{\includegraphics[width=18cm]{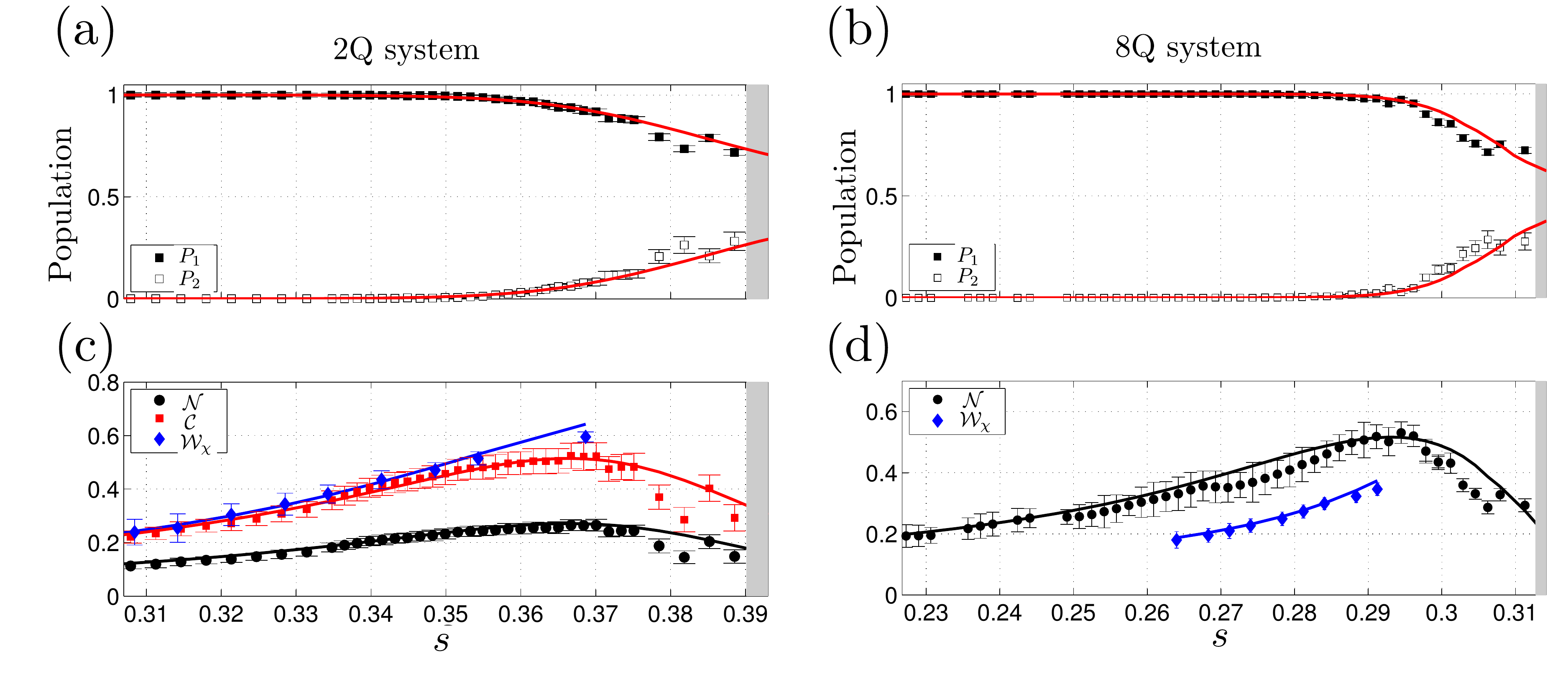}}
\caption{\label{FIG:entanglement-degree} {\bf{(a),(b)}} Measurements
  of the occupation fraction, or population, of the ground state
  ($P_1$) and first excited state ($P_2$) of the two-qubit and
  eight-qubit system, respectively, versus $s$. Early in the annealing
  trajectory, $g \gg k_B T$, and the system is in the ground state
  with $P_1 \lesssim 1$. The solid curves show the equilibrium
  Boltzmann predictions for $T = 12.5$ mK. {\bf (c)} Concurrence
  ${\cal C}$, negativity ${\cal N}$ and witness ${\cal W_{\chi}}$
  versus $s$ for the two-qubit system. Early in QA, the qubits are
  weakly interacting, thus resulting in limited
  entanglement. Entanglement peaks near $s = 0.37$.  For larger $s$,
  the gap between the ground and first excited state shrinks and
  thermal occupation of the first excited state rises, thus
  extinguishing entanglement. Solid curves indicate the expected
  theoretical values of each witness or measure using
  Hamiltonian~(\ref{HS}) and Boltzmann statistics. {\bf (d)}
  Negativity ${\cal N}$ and witness ${\cal W}_{\chi}$ versus $s$ for
  the eight-qubit system. For all $s$ shown, the nonzero negativity
  ${\cal N}$ and nonzero witness ${\cal W_{\chi}}$ report
  entanglement. For $s > 0.39$ and $s > 0.312$ for the two-qubit and
  eight-qubit systems, respectively, the shaded grey denotes the
  regime in which the ground and first excited states cannot be
  resolved via our spectroscopic method. Solid curves indicate the
  expected theoretical values of each witness or measure using
  Hamiltonian~(\ref{HS}) and Boltzmann statistics.}
\end{figure*}

The width of the measured spectral lines is dominated by the noise of
the probe device used to perform QTS~\cite{berkley2012}.  The probe
device was operated in a regime in which it is strongly coupled to its
environment, whereas the system qubits we studied are in the weak
coupling regime. The measured spectral widths therefore do not
represent the intrinsic width of the two- and eight-qubit energy
eigenstates. During the intermediate part of QA, the ground and
first excited states are clearly resolved. The ground state is
protected by the multi-qubit energy gap $g \gg k_BT$, and these
systems are coherent. At the end of the annealing trajectory, the gap
between the ground state and first excited state shrinks below the
probe qubit line width of 0.4 GHz. An analysis of the spectroscopy
data, which estimates the intrinsic level broadening of the
multi-qubit eigenstates, is presented in the Supplementary
Information. The analysis shows that the intrinsic energy levels
remain distinct until later in QA. The interactions between the two-
and eight-qubit systems and their respective environments represent
small perturbations to Hamiltonian~(\ref{HS}), even in the regime in
which entanglement is beginning to fall due to thermal mixing.

\section{VI. Entanglement Measures and Witnesses}

The tunneling spectroscopy data show that midway through QA, both the
two- and eight-qubit systems had avoided crossings with the expected
gap $g \gg k_BT$ and had ground state occupation $P_1\simeq 1$. While
observation of an avoided crossing is evidence for the presence of an
entangled ground state (see Ref.~\cite{WitPaper} and the Supplementary
Information for details), we can make this observation more
quantitative with entanglement measures and witnesses.

We begin with a susceptibility-based witness, ${\cal W}_{\chi}$, which
detects ground state entanglement. This witness does not require
explicit knowledge of Hamiltonian~(\ref{HS}), but requires a
non-degenerate ground state, confirmed with the avoided crossings
shown in Fig.~\ref{FIG:wide-spectrum-summary}, and high occupation
fraction of the ground state, confirmed early in QA by the
measurements of $P_1 \simeq 1$ shown in
Fig.~\ref{FIG:entanglement-degree}. We then performed measurements of
all available linear cross-susceptibilities $\chi_{ij} \equiv
d\left<\sigma_i^z\right>/d \tilde h_{j}$, where
$\left<\sigma_i^z\right>$ is the expectation value of $\sigma_i^z$ for
the $i$th qubit and $\tilde h_j = {\cal E} h_j$ is a bias applied to
the $j$th qubit. The measurements are performed at the degeneracy
point (in the middle of the avoided crossings) where the classical
contribution to the cross-susceptiblity is zero.

From these measurements, we calculated ${\cal W_{\chi}}$ as defined in
Ref.~\cite{WitPaper} (see Appendix D for more details). A non-zero
value of this witness detects ground-state entanglement, and global
entanglement in the case of the eight-qubit system (meaning every
possible bipartition of the eight-qubit system is entangled).
Figures~\ref{FIG:entanglement-degree}c and d show ${\cal W_{\chi}}$
for the two- and eight-qubit systems. Note that for two qubits at
degeneracy, ${\cal W_{\chi}}$ coincides with ground-state concurrence.
These results indicate that the two- and eight-qubit systems are
entangled midway through QA. Note also that a susceptibility-based
witness has a close analogy to susceptibility-based measurements of
nano-magnetic systems that also report strong non-classical
correlations~\cite{Vertesi06,Rappoport07}.

The occupation fraction measurements shown in
Fig.~\ref{FIG:entanglement-degree} indicate that midway through QA,
the first excited state of these systems is occupied as the energy gap
$g$ begins to approach $k_BT$. The systems are no longer in the ground
state, but, rather, in a mixed state.  To detect the presence of mixed-state
entanglement, we need knowledge about the density matrix of these
systems. Occupation fraction measurements provide measurements of the
diagonal elements of the density matrix in the energy basis.  We assume
that the density matrix has no off-diagonal elements in the energy
basis (they decay on timescales of several ns). We relax this
assumption below. Populations $P_1$ and $P_2$ plotted in
Figs.~\ref{FIG:entanglement-degree}a and
~\ref{FIG:entanglement-degree}b indicate that the system occupies
these states with almost 100\% probability.  This means that the
density matrix can be written in the form $\rho = \sum_{i=1}^2 P_i
\ket{\psi_i}\bra{\psi_i}$ where $\ket{\psi_i}$ represents the $i$th
eigenstate of Hamiltonian~(\ref{HS}).

We use the density matrix to calculate standard entanglement measures,
Wootters' concurrence, ${\cal C}$~\cite{wootters1998}, for the
two-qubit system, and negativity, ${\cal N}$~\cite{Guhne09,VW02}, for
the two- and eight-qubit system. For the maximally entangled two-qubit
Bell state we note that ${\cal C} = 1$ and ${\cal N} = 0.5$. Figure
\ref{FIG:entanglement-degree}c shows ${\cal C}$ as a function of
$s$. Midway through QA we measure a peak concurrence ${\cal C} =
0.53\pm0.05$, indicating significant entanglement in the two-qubit
system. This value of ${\cal C}$ corresponds to an entanglement of
formation $E_f = 0.388$ (see Refs. \cite{Guhne09,wootters1998} for
definitions).  This is comparable to the level of entanglement, $E_f =
0.378$, obtained in Ref.~\cite{ansmann}, and indeed to the value
$E_f=1$ for the Bell state. Because concurrence ${\cal C}$ is not
applicable to more than two qubits, we used negativity ${\cal N}$ to
detect entanglement in the eight-qubit system. For $N>2$, ${\cal
  N}_{A,B}$ is defined on a particular bipartition of the system into
subsystems $A$ and $B$. We define ${\cal N}$ to be the geometric mean
of this quantity across all possible bipartitions. A nonzero ${\cal
  N}$ indicates the presence of global entanglement.
Figures~\ref{FIG:entanglement-degree}c and d show the negativity
calculated with measured $P_1$ and $P_2$ (and with the measured
Hamiltonian parameters $\Delta$ and ${\cal E}J_{ij}$) as a function of
$s$ for the two and eight-qubit systems. The eight-qubit system has
nonzero ${\cal N}$ for $s < 0.315$, thus indicating the presence of
mixed-state global entanglement. Both concurrence ${\cal C}$ and
negativity ${\cal N}$ decrease later in QA where the first excited
state approaches the ground state and becomes thermally occupied. The
experimental values of these entanglement measures are in agreement
with the theoretical predictions (solid lines in
Fig~\ref{FIG:entanglement-degree}). The error bars in
Figures~\ref{FIG:entanglement-degree}c and d represent uncertainties
in the measurements of occupation fractions, $\Delta(s)$ and ${\cal
  E}(s)$.

\begin{figure}
\begin{center}
\includegraphics[trim=4cm 6.5cm 2.5cm
  6.5cm,clip,width=8.5cm]{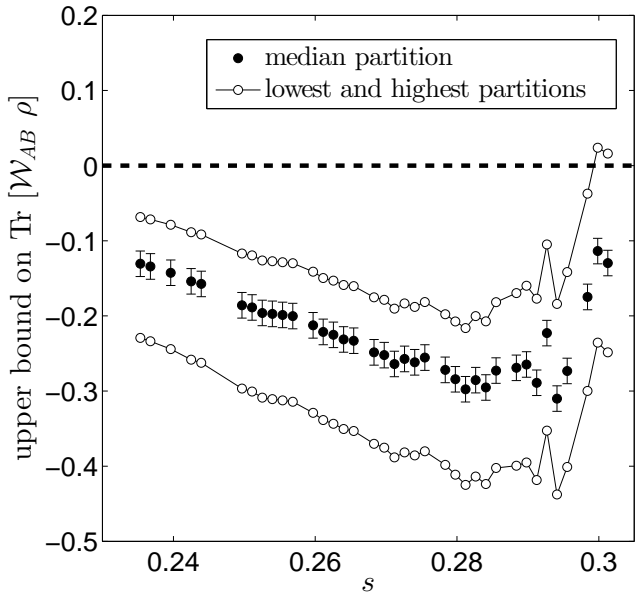}
\end{center}
\caption{\label{FIG:witness-a} Upper limit of the quantity Tr$[{\cal
      W}_{AB}\rho]$ versus $s$ for several bipartitions $A-B$ of the
  eight-qubit system. When this quantity is $<0$, the system is
  entangled with respect to this bipartition. The solid dots show the
  upper limit on Tr$[{\cal W}_{AB}\rho]$ for the median
  bipartition. The open dots above and below these are derived from
  the two bipartitions that give the highest and lowest upper limits
  on Tr$[{\cal W}_{AB}\rho]$, respectively.  For the points at
  $s>0.3$, the measurements of $P_1$ and $P_2$ do not constrain $\rho$
  enough to certify entanglement.}

\end{figure}

As stated above, the calculation of $\cal{C}$ and ${\cal N}$ relies on
the assumption that the off-diagonal terms in the density matrix decay
on times scales of several ns. We remove this assumption and
demonstrate entanglement through the use of another witness ${\cal
  W}_{AB}$, defined on some bipartition $A$-$B$ of the eight-qubit
system. The witness, described in Appendix D, is designed in such a
way that Tr$[{\cal W}_{AB}\sigma] \geq 0 $ for all separable
states $\sigma$. When Tr$[ {\cal W}_{AB}\rho(s)]<0$, the state $\rho(s)$ is
entangled.  Measurements of populations $P_1$ and $P_2$ provide a set
of linear constraints on the density matrix of the system,
$\rho(s)$. We then obtain an upper bound on Tr$[{\cal
    W}_{AB}\rho(s)]$ by searching over all $\rho(s)$ that satisfy these
linear constraints. If this upper bound is $<0$, then we have shown
entanglement for the bipartition
$A$-$B$~\cite{spedalieri2012}. Figure~\ref{FIG:witness-a} shows the
upper limit of the witness Tr$[ {\cal W}_{AB}\rho(s)]$ for the
eight-qubit system.  We plot data for the bipartition that gives the
median upper limit. The error bars are derived from a Monte-Carlo
analysis wherein we used the experimental uncertainties in $\Delta$
and $J$ to estimate the uncertainty in Tr$[{\cal W}_{AB}\rho]$. We
also plot data for the two partitions that give the largest and
smallest upper limits. For all values of the annealing parameter $s$,
except for the last two points, upper limits from all possible
bipartitions of the eight-qubit system are below zero. In this
annealing range, the eight-qubit system is globally entangled.

\section{VII. Conclusions}

To summarize, we have provided experimental evidence for the presence
of quantum coherence and entanglement within subsets of qubits inside
a quantum annealing processor during its operation.  Our conclusion is
based on four levels of evidence: \textbf{a.}~the observation of two-
and eight- qubit avoided crossings with a multi-qubit energy gap $g
\gg k_B T$; ~\textbf{b.}~the witness ${\cal W}\chi$, calculated with
measured cross-susceptibilities and coupling energies, which reports
ground state entanglement of the two- and eight-qubit system. Notice
that these two levels of evidence do not require explicit knowledge of
Hamiltonian~(\ref{HS}); ~\textbf{c.}~the measurements of energy
eigenspectra and equibrium occupation fractions during QA, which allow
us to use Hamiltonian~(\ref{HS}) to reconstruct the density matrix,
with some weak assumptions, and calculate concurrence and
negativity. These standard measures of entanglement report
non-classical correlations in the two- and eight-qubit systems;
\textbf{d.}~the entanglement witness ${\cal W}_{AB}$, which is
calculated with the measured Hamiltonian and with constraints provided
by the measured populations of the ground and the first excited
states. This witness reports global entanglement of the eight-qubit
system midway through the QA algorithm.

The observed entanglement is persistent at thermal equilibrium, an
encouraging result as any practical hardware designed to run a quantum
algorithm will be inevitably coupled to a thermal environment.  The
experimental techniques that we have discussed provide measurements of
energy levels, and their populations, for arbitrary configurations of
Hamiltonian parameters $\Delta, h_i, J_{ij}$ during the QA
algorithm. The main limitation of the technique is the spectral width
of the probe device. Improved designs of this device will allow much
larger systems to be studied.  Our measurements represent an effective
approach for exploring the role of quantum mechanics in QA processors
and ultimately to understanding the fundamental power and capability
of quantum annealing.

\section{Acknowledgements}
We thank C.~Williams, P.~Love, and J.~Whittaker for useful
discussions. We acknowledge F.~Cioata and P.~Spear for the design and
maintenance of electronics control systems, J.~Yao for fabrication
support, and D.~Bruce, P.~deBuen, M.~Gullen, M.~Hager, G.~Lamont,
L.~Paulson, C.~Petroff, and A.~Tcaciuc for technical
support. F.M.S. was supported by DARPA, under contract
FA8750-13-2-0035.


\appendix
\section{Appendix A. QA Processor Description}
\subsection{Chip Description}
The experiments discussed in herein were performed on a sample
fabricated with a process consisting of a standard Nb/AlOx/Nb
trilayer, a TiPt resistor layer, planarized SiO$_2$ dielectric layers
and six Nb wiring layers. The circuit design rules included a minimum
linewidth of 0.25 $\mu$m and $0.6\ \mu$m diameter Josephson
junctions. The processor chip is a network of densely connected
eight-qubit unit cells which are more sparsely connected to each other
(see Fig.~1 for photographs of the processor). We report measurements
made on qubits from one of these unit cells. The chip was mounted on
the mixing chamber of a dilution refrigerator inside an Al
superconducting shield and temperature controlled at 12.5 mK.

\subsection{Qubit Parameters}

The processor facilitates quantum annealing (QA) of compound-compound
Josephson junction rf SQUID (radio-frequency superconducting quantum
interference device) flux qubits~\cite{Harris-CCJJ-2010}. The qubits
are controlled via the external flux biases $\Phi_{qi}^x$ and
$\Phi_{\rm ccjj}^x$ which allow us to treat them as effective spins
(see Fig.~1). Pairs of qubits interact through tunable inductive
couplings~\cite{harris-cjc-2009}. The system can be described with the
time-dependent QA Hamiltonian,
 \begin{equation}
 {\cal H}_S(s) = {\cal E}(s) \left[ - \sum^N_{i} h_i \sigma^z_i +
   \sum_{i<j} J_{ij}\sigma^z_i\sigma^z_j\right] -\frac{1}{2}\,
 \Delta(s)\, \sum^N_{i} \sigma^x_i, \label{HSS}
 \end{equation}
where $\sigma_i^{x,z}$ are Pauli matrices for the $i$th qubit, $i = 1,
\ldots,N.$ The energy scales $\Delta$ and ${\cal E}$ are the
transverse and longitudinal energies of the spins, respectively, and
the unitless biases $h_i$ and couplings $J_{ij}$ encode a particular
optimization problem. We define $\tilde{h_i} \equiv {\cal E}h_i$ and
$\tilde{J_{ij}} \equiv {\cal E}J_{ij}$. We have mapped the annealing
parameter $s$ for this particular chip to a range of $\Phi_{\rm
  ccjj}^x$ with the relation
\begin{equation}
s \equiv (\Phi_{\rm ccjj}^x(t) - \Phi_{{\rm ccjj},{\rm
    initial}}^x)/(\Phi_{{\rm ccjj},{\rm final}}^x - \Phi_{{\rm
    ccjj},{\rm initial}}^x) = t/t_f,
\end{equation}
where $t_f$ is the total anneal time. We implement QA for this
processor by ramping the external control $\Phi_{\rm ccjj}^x(t)$ from
$\Phi_{\rm{ccjj},{\rm initial}}^x = 0.596\ \Phi_0\ (s=0)$ at $t=0$ to
$\Phi_{\rm{ccjj},{\rm final}}^x = 0.666\ \Phi_0\ (s=1)$ at
$t=t_f$. The energy scale ${\cal E} \equiv M_{\rm eff}|I_q^p(s)|^2$ is
set by the $s$-dependent persistent current of the qubit $|I_q^p(s)|$
and the maximum mutual inductance between qubits $M_{\rm eff} = 1.37$
pH~\cite{harris-2010}. The transverse term in Hamiltonian (\ref{HSS}),
$\Delta(s)$, is the energy gap between the ground and first excited
state of an isolated rf SQUID at zero bias. $\Delta$ also changes with
annealing parameter $s$. $\Phi_{qi}^x(t)$ is provided by a global
external magnetic flux bias along with local {\em in situ} tunable
digital-to-analog converters (DAC) that tune the coupling strength of
this global bias into individual qubits and thus allow us to specify
individual biases $h_i$. The coupling energy between the $i$th and
$j$th qubit is set with a local {\em in situ} tunable DAC that
controls $\Phi_{{\rm co},ij}^x$.

\begin{figure}[ht]
\includegraphics[trim=1cm 6cm 2cm 6cm,clip,width=9cm]{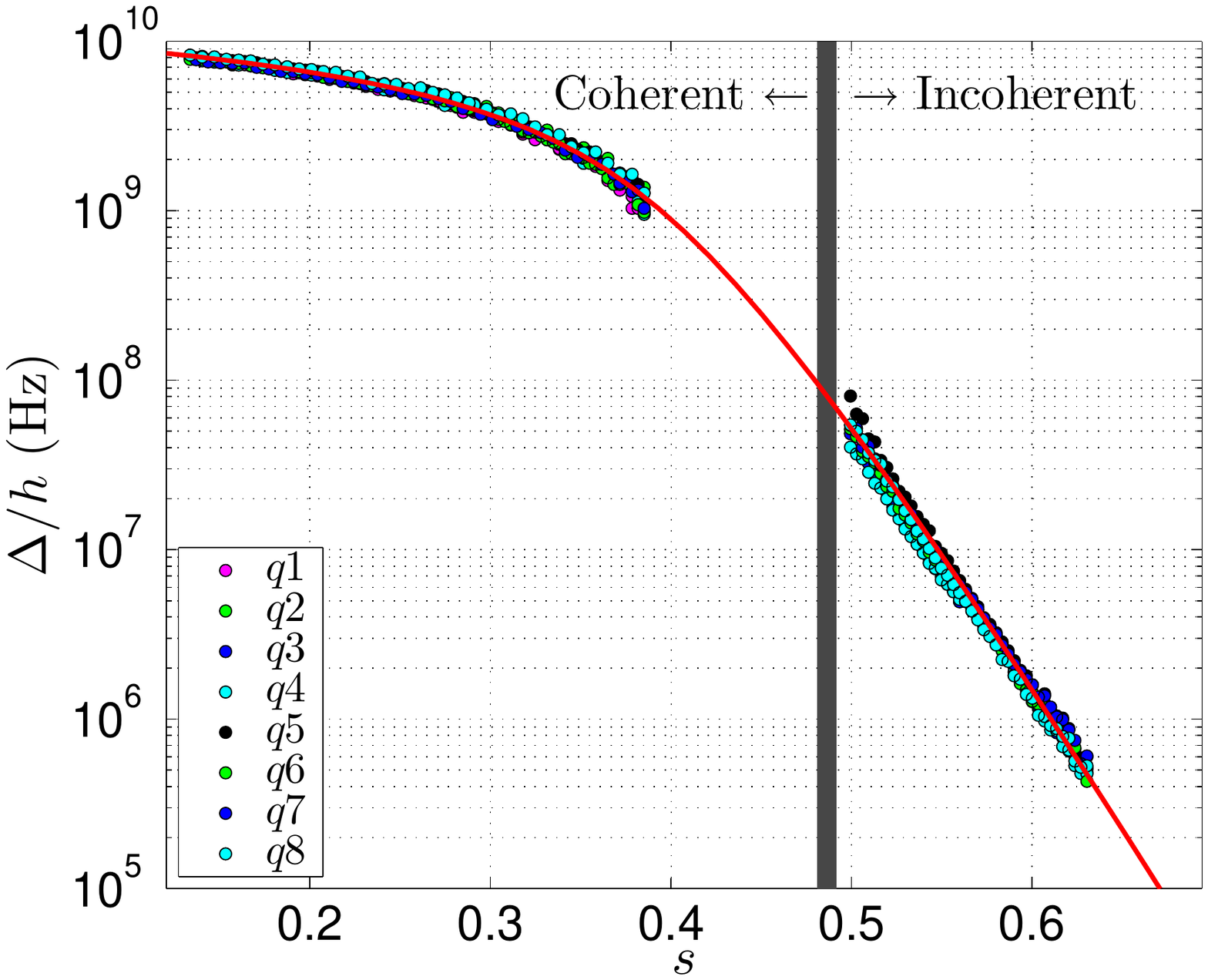}\\
\includegraphics[trim=1cm 6cm 2cm 6cm,clip,width=9cm]{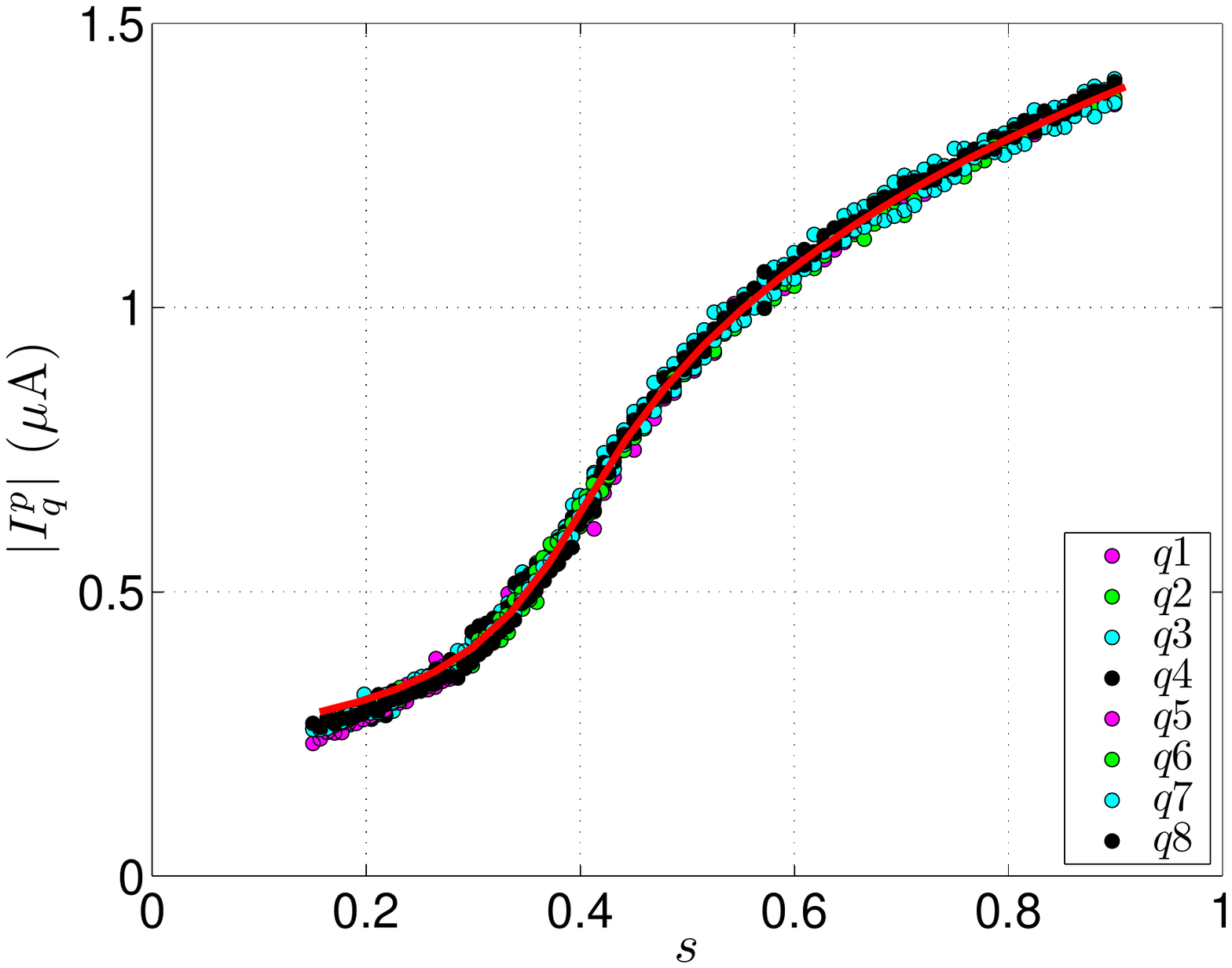}
\caption{({\bf a}) $\Delta(s)$ vs $s$. We show measurements for all
eight qubits studied in this work. We used a single qubit
Landau-Zener experiment to measure $\Delta/h < 100$ MHz~\cite{Johansson2009}. We used
qubit tunneling spectroscopy (QTS) to measure $\Delta/h > 1$ GHz~\cite{berkley2012}. The red line shows the
theoretical prediction for an rf SQUID model employing the median
qubit parameters of the eight devices. The vertical black line
separates coherent (left) and incoherent (right) evolution as estimated
by analysis of single qubit spectral line shapes. ({\bf b})
$|I_q^p|(s)$ vs $s$. We show measurements for all eight qubits
studied in this work. We used a two-qubit coupled flux measurement
with the inter-qubit coupling element set to $1.37$ pH~\cite{harris-2010}. The red line
shows the theoretical prediction for an rf SQUID model employing the
median qubit parameters of the eight devices.
\label{FIG:delta-vs-s}}
\end{figure}

The main quantities associated with a flux qubit, $\Delta$ and
$|I_q^p|$, primarily depend on macroscopic rf SQUID parameters:
junction critical current $I_c$, qubit inductance $L_q$, and qubit
capacitance $C_q$. We calibrated all of these parameters on this
chip as described in \cite{NaturePaper,harris-2010}. We calibrated all
inter-qubit coupling elements across their available tuning range
from $1.37$ pH to $-3.7$ pH as described in Ref.~\cite{harris-cjc-2009}. We corrected for variations in qubit
parameters with on-chip control as described in \cite{harris-2010}.
This allowed us to match $|I_q^p|$ and $\Delta$ across all qubits
throughout the annealing trajectory. Table \ref{TBL:qubitparameters}
shows the median qubit parameters for the devices studied here.

\begin{table}
\begin{tabular}{|c|c|}
\hline
qubit parameter & median measured value \\
\hline
 critical current, $I_c$ & 2.89 
 $\mu$A \\
qubit inductance, $L_q$ & 344 
pH \\
qubit capacitance, $C_q$ & $110 
$ fF \\
\hline
\end{tabular}
\caption{Qubit Parameters. \label{TBL:qubitparameters}}
\end{table}

Figure \ref{FIG:delta-vs-s} shows measurements of $\Delta$ and
$|I_q^p|$ vs. $s$ for all eight qubits. $\Delta$ was measured with
single qubit Landau-Zener measurements from $s = 0.515$ to $s =
0.658$~\cite{Johansson2009} and with qubit tunneling spectroscopy
(QTS) from $s = 0.121$ to $s = 0.407$~\cite{berkley2012}. The
resolution limit of qubit tunneling spectroscopy and the bandwidth of
our external control lines during the Landau-Zener measurements
prevented us from characterizing $\Delta$ between $s= 0.4$ and
$s=0.5$, respectively. $|I_q^p|$ was measured by coupling a second
probe qubit to the qubit $q_i$ with a coupling of $M_{\rm eff} = 1.37$
pH and measuring the the flux $M_{\rm eff}|I_{qi}^p(s)|$ as a function
of $s$. $|I_q^p|$ is matched between qubits to within $3\%$ and
$\Delta(s)$ is matched between qubits to within $8\%$ across the
annealing region explored in this study.

\section{Appendix B. Qubit Tunneling Spectroscopy (QTS)}

QTS allows one to measure the eigenspectrum of an $N$-qubit system
governed by Hamiltonian ${\cal H}_S$. Details on the measurement
technique are presented elsewhere~\cite{berkley2012}. For convenience
in comparing with this reference, we define a qubit energy bias
$\epsilon_i \equiv 2\tilde{h_i}$. Measurements are performed by
coupling an additional probe qubit $q_P$, with qubit tunneling
amplitude $\Delta_P \ll \Delta, |\tilde{J}|$, to one of the $N$ qubits
of the system under study, for example $q_1$. When we use a coupling
strength $\tilde J_P$ between $q_P$ and $q_1$ and apply a compensating
bias $\epsilon_1 = 2\tilde{J_P}$ to $q_1$, the resulting \emph{system
  + probe} Hamiltonian becomes

 \begin{equation}
 H_{S+P} = H_S - [ \tilde{J_P} \sigma^z_1 - (1/2)\, \epsilon_P ]
 (1-\sigma^z_P). \label{HSP}
 \end{equation}

For one of the localized states of the probe qubit, $\ket{\uparrow}_P
$, for which an eigenvalue of $\sigma^z_P$ is equal to +1 (i.e. the
probe qubit in the right well), the contribution of the probe qubit is
exactly canceled, leading to $H_{S+P}=H_S$, with composite eigenstates
$\ket{n,\uparrow} = \ket{n}\otimes\ket{\uparrow}_P$ and eigenvalues
$E_n^R = E_n$, which are identical to those of the original system
without the presence of the probe qubit. Here, $\ket{n}$ is an
eigenstate of the Hamiltonian $H_S$ $(n = 1,2,...,2^N)$.

For the other localized state of the probe qubit,
$\ket{\downarrow}_P$, when this qubit is in the left well, the ground
state of $H_{S+P}$ is $\ket{\psi^L_0,\downarrow} =
\ket{\psi^L_0}\otimes\ket{\downarrow}_P$, with eigenvalue $\widetilde
E_0^L = E_0^L + \epsilon_P$, where $\ket{\psi_0^L}$ is the ground
state of $H_S- 2\tilde{J_P}\sigma^z_1$ and $E_0^L$ is its
eigenvalue. We choose $|\tilde{J_P}| \gg k_BT$ such that the state
$\ket{\psi_0^L,\downarrow}$ is well separated from the next excited
state for ferromagnetically coupled systems, and thus \emph{system + probe} can be initialized in this state
to high fidelity.

\begin{figure}[ht]
\includegraphics[width=0.5\textwidth]{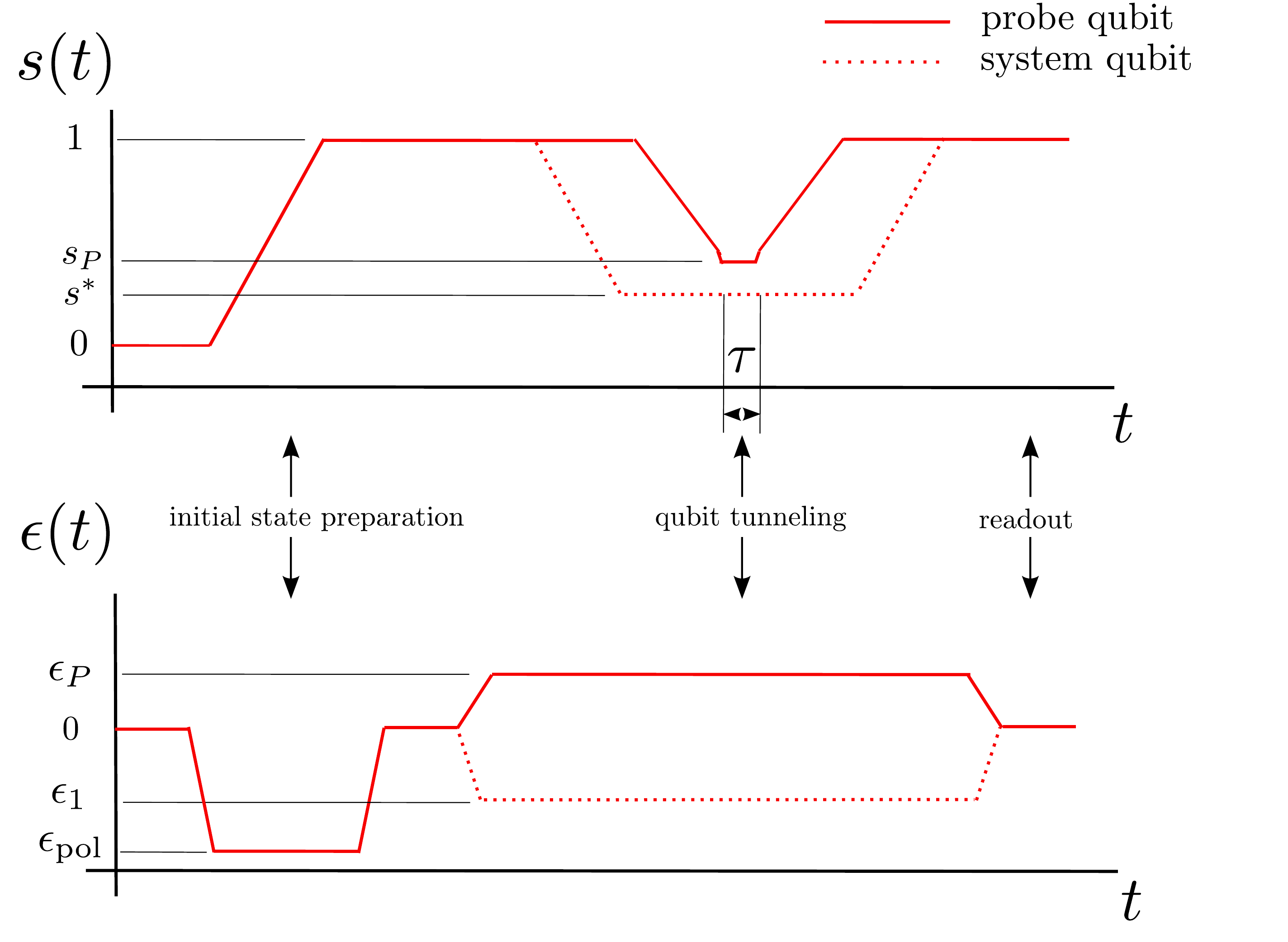}
\caption{\label{FIG:pulse-sequence} Typical waveforms during QTS. We
  prepare the initial state by annealing probe and system qubits from
  $s=0$ to $s=1$ in the presence of a large polarization bias
  $\epsilon_{\rm pol}$. We then bias the system qubit $q_1$ (to which
  the probe is attached) to a bias $\epsilon_1$ and the probe qubit to
  a bias $\epsilon_P$. With these biases asserted, we then adjust the
  system qubits' annealing parameter to an intermediate point $s^*$
  and the probe qubit to a point $s_P$ and dwell for a time
  $\tau$. Finally, we complete the anneal $s\rightarrow 1$ and read
  out the state of the qubits.} \end{figure}

Introducing a small transverse term, $-{1\over2}\Delta_P\sigma^x_P$,
to Hamiltonian (\ref{HSP}) results in incoherent tunneling from the
initial state $\ket{\psi_0^L,\downarrow}$ to any of the available
$\ket{n,\uparrow}$ states~\cite{harris-mrt-2008}. A bias
on the probe qubit, $\epsilon_P$, changes the energy difference
between the probe $\ket{\downarrow}_P$ and $\ket{\uparrow}_P$
manifolds. We can thus bring $\ket{\psi_0^L,\downarrow}$ into
resonance with any of $\ket{n,\uparrow}$ states (when $\widetilde
E_0^L = E_n^R$) allowing resonant tunneling between the two states.
The rate of tunneling out of the initially prepared state
$\ket{\psi_0^L,\downarrow}$ is thus peaked at the locations of
$\ket{n,\uparrow}$.

The measurement of the eigenspectrum of an $N$-qubit system thus
proceeds as follows. We couple an additional probe qubit to one of
the $N$-qubits (say, to $q_1$) with coupling constant $\tilde{J_P}$.
We prepare the $N$+1-qubit system in the state
$\ket{\psi^L_0,\downarrow}$ by annealing from $s = 0$ to $s = 1$ in
the presence of large bias $\epsilon_{\rm pol} < 0$ on all the
system and probe qubits. We then adjust $s$ for the $N$-qubit system
to an intermediate point $s^* \in [0, 1]$ such that $\Delta \gg
k_BT/h$ and $s$ for the probe qubit to $s_P = 0.612$ such that
$\Delta_P/h \sim 1$ MHz (here $h$ is the Planck constant).  We
assert a compensating bias $\epsilon_1 = 2\tilde{J_P}$ to this
qubit. We dwell at this point for a time $\tau$, complete the anneal
$s \rightarrow 1$ for the {\em system+probe}, and then read out the
state of the probe qubit. Figure \ref{FIG:pulse-sequence} summarizes
these waveforms during a typical QTS
measurement.

We perform this measurement for a range of $\tau$ which allows us to
measure an initial rate of tunneling $\Gamma$ from
$\ket{\psi^L_0,\downarrow}$ to $\ket{\psi,\uparrow}$. We repeat this
measurement of $\Gamma$ for a range of the probe qubit bias
$\epsilon_P$. Peaks in $\Gamma$ correspond to resonances between the
initially prepared state and the state $\ket{n,\uparrow}$, thus
allowing us to map the eigenspectrum of the $N$-qubit system.

\begin{figure*}[t]
centering \setlength\fboxsep{0pt} \setlength\fboxrule{0pt}
\fbox{\includegraphics[width=18cm]{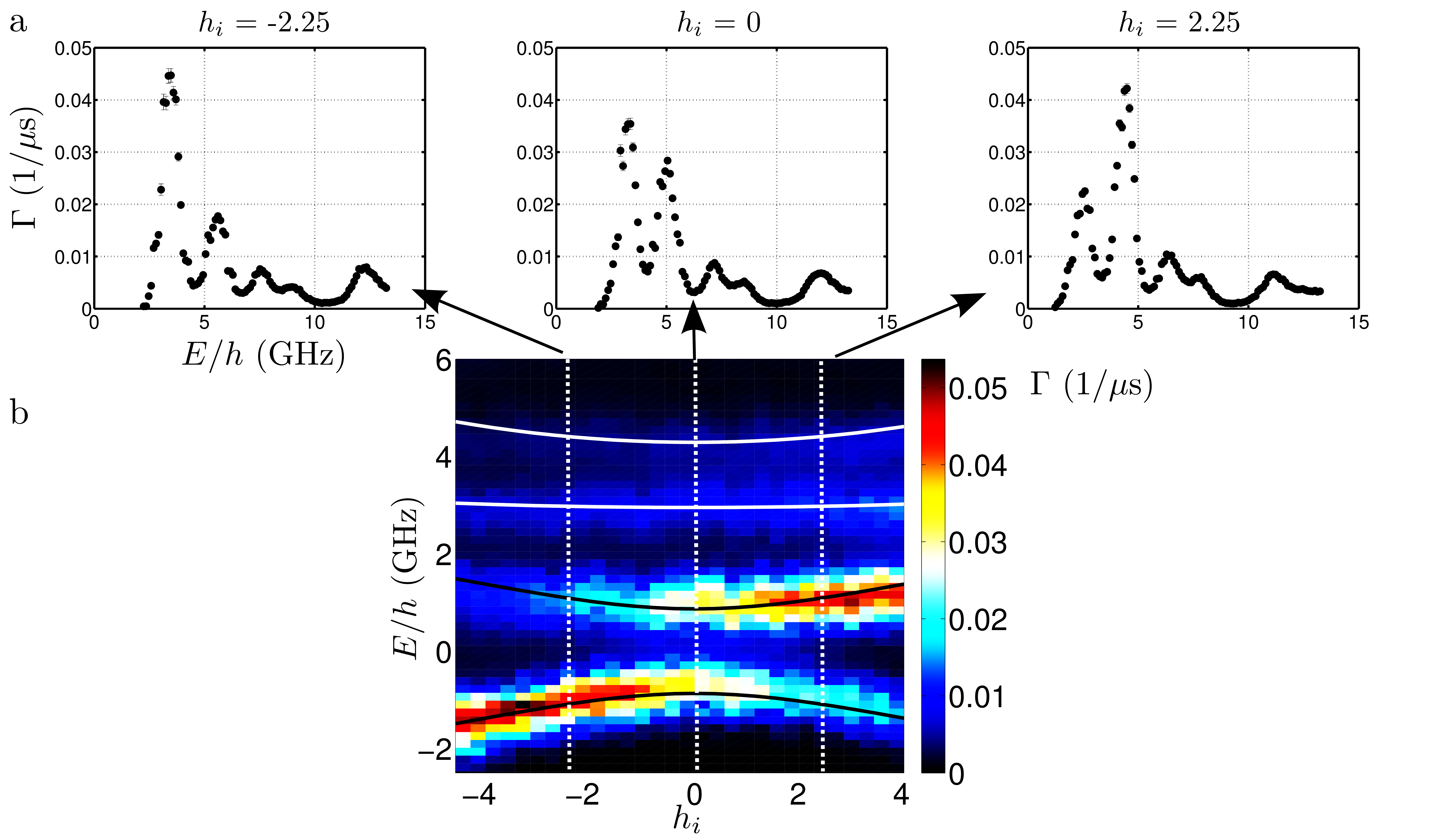}}
\caption{\label{FIG:2q-with-raw} Spectroscopy data for two FM coupled
  qubits at $\tilde{J_P}<0$. ({\bf a}) Measurements of tunneling rate
  $\Gamma$ for three values of $h_1 = h_2\equiv h_i$.  These data were
  taken at $s = 0.339$. Peaks in $\Gamma$ reveal the energy
  eigenstates of the two-qubit system. ({\bf b}) Multiple scans of
  $\Gamma$ for different values of $h_i$ assembled into a
  two-dimensional color plot. For better interpretability, we have
  subtracted off a baseline energy with respect to ({\bf a}) such that
  the ground and first excited levels are symmetric about zero. Notice
  the avoided crossing at $h_i = 0$. The peak tunneling rate $\Gamma
  \sim |\Delta_P\bra{\psi_0^L}n\rangle|^2$~\cite{berkley2012}. The
  solid black and white curves plot the theoretical expectations for
  the energy eigenvalues using independent measurements shown in
  Figure~\ref{FIG:delta-vs-s} and Hamiltonian~(\ref{HS}).}
\end{figure*}

For the plots in the main paper, measurements of $\Gamma$ are
normalized to $[0,1]$ by dividing the maximum value across a
vertical slice to give a visually interpretable result. Figure
\ref{FIG:2q-with-raw}b shows a typical raw result in units of $\mu
s^{-1}$.

We posed ferromagnetically coupled instances of the form
\begin{equation}
{\cal H}_P = - \sum_{i}h_i\sigma^z_i + \sum_{i<j}
J_{ij}\sigma^z_i\sigma^z_j \label{HP}
\end{equation}
with $J_{ij} < 0$ for two and eight qubit subsections of the QA
processor. Figure~\ref{FIG:2q-with-raw}a shows typical measurements of
$\Gamma$ for a two qubit subsection at several biases $h_i$ and at $s
= 0.339$ ($\tilde{J_P}<0$). We assembled multiple measurements to
produce the spectrum shown in Figure~\ref{FIG:2q-with-raw}b.

\section{Appendix C. Equilibrium Distribution of System}

In addition to the energy eigenspectrum, QTS also provides a means of
measuring the equilibrium distribution of an $N$-qubit system with a
probe qubit. Suppose we are in the limit $|\tilde{J_P}| \gg k_B T$
such that there is only one accessible state in the
$\ket{\downarrow}_P$ manifold:
$\ket{\psi_0^L}\otimes\ket{\downarrow}_P$. As described above, the
other available states in the system are the composite eigenstates
$\ket{n}\otimes\ket{\uparrow}_P$ in the $\ket{\uparrow}_P$ manifold
where $\ket{n}$ is an eigenstate of the $N$-qubit system without the
probe qubit attached. Energy levels $E_n^R$ of the $\ket{\uparrow}_P$
manifold coincide with the energy levels $E_n$ of the system, $E_n^R =
E_n$, even in the presence of coupling between the probe qubit and the
system. We make the assumption that the population of an eigenstate
depends only on its energy. Degenerate states have the same
population.

Let $P^L$ represent the probability of finding the {\em probe+system}
in the state $\ket{\psi_0^L}\otimes\ket{\downarrow}_P$ and $P^R_n$
represent the probability of finding the {\em probe+system} in the
state $\ket{n}\otimes\ket{\uparrow}_P$. At any point in the {\em
  probe+system} evolution we expect:

\begin{equation}
P^L + \sum_{i=1}^{2^N}P^R_i = 1 \label{EQN:sumtounity}
\end{equation}

As described in the previous section, we can alter the energy of
$\ket{\psi_0^L}\otimes\ket{\downarrow}_P$ with the probe bias
$\epsilon_P$. Based on the spectroscopic measurements of the $N$-qubit
eigenspectrum, we can choose an $\epsilon_P$ such that
$\ket{\psi_0^L}\otimes\ket{\downarrow}_P$ and
$\ket{n}\otimes\ket{\uparrow}_P$ are degenerate. Since the
occupation of the state depends on its energy, we expect that, after
long evolution times, these two degenerate states are occupied with
equal probability, $P^L(\epsilon_P{=}E_n) = P^R_n$. Aligning the
state $\ket{\psi_0^L}\otimes\ket{\downarrow}_P$ with all possible
$2^N$ states $\ket{n}\otimes\ket{\uparrow}_P$ we obtain a set of
relative probabilities $P_n^R$. These relative probabilities
characterize the population distribution in the system since they
are uniquely determined by the energy spectrum $E_n$. However, as
follows from Eq.~(\ref{EQN:sumtounity}), the set $P_n^R$ is not
properly normalized. The probability distribution of the system
itself is given by:
\begin{equation}
P_n(E_n) = \frac{P^R_n}{\sum_{i=1}^{2^N}P^R_i}, \label{EQN:PNN}
\end{equation}
where $\sum_{n=1}^{2^N} P_n = 1$. At every eigenenergy, $\epsilon_P = E_n$, the denominator of
Eq.~(\ref{EQN:PNN}) can be found from Eq.~(\ref{EQN:sumtounity}), so
that the population distribution of the system $P_n$ has the form
\begin{equation}
P_n =   \frac{P^R_n}{1-P^L} =
\frac{P^L(\epsilon_P{=}E_n)}{1-P^L(\epsilon_P{=}E_n)}.
\label{EQN:PN}
\end{equation}
Thus, the probability $P_n$ to find the system of $N$ qubits
in the state with energy $E_n$ can be estimated by measuring $P^L$
at $\epsilon_P = E_n$ and using Equation (\ref{EQN:PN}).


Measurements of $P^L$ proceed as they do for the spectroscopy
measurements. The {\em system+probe} is prepared in
$\ket{\psi_0^L,\downarrow}$. We then adjust $\epsilon_P = E_n$, and
an annealing parameter $s$ for the $N$-qubit system to some
intermediate point, and also $s_P = 0.612$ for the probe qubit such
that $\Delta_{P}/h \sim 1$ MHz. We dwell at this point for a time
$\tau \gg 1/\Gamma$, complete the anneal $s\rightarrow 1$, and then
read out the state of the probe qubit. We typically investigate a
range of $\tau$ to ensure that we are in the long evolution time
limit in which $P^L$ is independent of $\tau$. We use $P^L$ measured
with $\tau = 7041$ $\mu$s to estimate $P_1$ and $P_2$. The
Supplementary Information contains typical data used for these
estimates.

\section{Appendix D.}
\subsection{Susceptibility-based entanglement witness
${\cal W}_{\chi}$}

For a bipartion of the system into two parts, $A$ and $B$, we define a
witness ${\cal R}_{AB}$ as
\begin{equation} \label{RAB}
{\cal R}_{AB}  = \frac{1}{ 4 N_{AB}} \;|\sum_{i\in A} \sum_{j\in
B}\, \tilde J_{ij}\, \chi_{ij}|,
\end{equation}
where $\chi_{ij}$ is a cross-susceptibility, $\tilde J_{ij} = {\cal
E} J_{ij}$, and $N_{AB}$ is a number of non-zero couplings, $J_{ij}
\neq 0$, between qubits from the subset $A$ and the subset $B$ (see
Ref.~\cite{WitPaper} and the Supplementary Information). We note
that at low temperature, $T = 12.5$~mK, the measured susceptibility
$\chi_{ij}(T)$ almost coincides with the ground-state susceptibility
$\chi_{ij}(T=0)$ since contributions of excited states to
$\chi_{ij}(T)$ are proportional to their populations, $P_n \ll 1$,
for $n > 1$. We analyze a deviation of the measured susceptibility
from its ground-state value in the Supplementary Information. To
characterize global entanglement in the system of $N$ qubits we
introduce a witness ${\cal W}_{\chi}$,
\begin{equation} \label{Wchi}
{\cal W}_{\chi} = \sqrt{\frac{\left(\prod {\cal
R}_{AB}\right)^{1/N_p}}{ 1 + \left(\prod {\cal
R}_{AB}\right)^{1/N_p} }},
\end{equation}
which is given by a bounded geometrical mean of witnesses ${\cal
R}_{AB}$ calculated for all possible partitions of the whole system
into two subsystems. Here $N_p$ is a number of such bipartitions, in
particular, $N_p = 127$ for the eight-qubit ring.

\subsection{Entanglement witness ${\cal W}_{AB}$}

Consider Hamiltonian~(\ref{HS}) with measured parameters. This
Hamiltonian describes a transverse Ising model having $N$ qubits.  The
ground state $|\psi_1\rangle$ of this model is entangled with respect
to some bipartition $A-B$ of the $N$-qubit system. We can form an
operator $|\psi_1\rangle\langle\psi_1|^{T_{A}}$ where $T_A$ is a
partial transposition operator with respect to the $A-$subsystem
\cite{Guhne09}. Let $|\phi\rangle$ be the eigenstate of
$|\psi_1\rangle\langle\psi_1|^{T_{A}}$ with the most negative
eigenvalue. We can form a new operator ${\cal W}_{AB} =
|\phi\rangle\langle\phi|^{T_{A}}$. This operator can serve as an
entanglement witness (it is trivially positive on all separable
states).

Let $\rho(s)$ be the density matrix associated with the state of the
system at the annealing point $s$. If we have experimental
measurements of the occupation fraction of the ground state and first
excited state, $P_1(s)\pm \delta P_1$ and $P_2(s)\pm \delta P_2$,
respectively, we can place a set of linear constraints on $\rho(s)$:

 \begin{eqnarray}
 \mathrm{Tr}[\rho(s) |\psi_1 \rangle\langle \psi_1|] &\geq
 P_1(s)-\delta P_1 \nonumber \\ \mathrm{Tr}[\rho(s) |\psi_1
   \rangle\langle \psi_1|] &\leq P_1(s)+\delta P_1 \nonumber
 \\ \mathrm{Tr}[\rho(s) |\psi_2 \rangle\langle \psi_2|] &\geq
 P_2(s)-\delta P_2 \nonumber \\ \mathrm{Tr}[\rho(s) |\psi_2
   \rangle\langle \psi_2|] &\leq P_2(s)+\delta P_2 \nonumber
 \end{eqnarray}

We now search over all possible $\rho(s)$ that satisfy the linear
constraints provided by the experimental data. The goal is to maximize
the witness Tr$[{\cal W}_{AB}\rho(s)]$ in order to establish an upper
limit for this quantity. Maximizing this quantity can be cast as a
semidefinite program \cite{spedalieri2012}, a class of convex
optimization problems for which efficient algorithms exist. When this
upper limit is less than zero, entanglement is certified for the
bipartition $A-B$.

We tested the robustness of this result with uncertainties in the
parameters of the Hamiltonian. To do this, we have repeated the
analysis at several points during the QA algorithm when adding
random perturbations on the measured Hamiltonian that correspond to
the uncertainty on these measured quantities. We sampled $10^4$
perturbed Hamiltonians and, for every perturbation, the optimization
resulted in Tr$[{\cal W}_{AB}\rho(s)] < 0$.

\end{document}